\documentclass{cta-author}

\usepackage{float}
\usepackage{amsmath,graphicx,amssymb,algorithm}
\usepackage{flushend}
\usepackage{xspace}
\usepackage{epstopdf}
\usepackage{tikz}
\usepackage{tikz-3dplot}
\usetikzlibrary{shapes.geometric,arrows}
\usetikzlibrary{shapes.misc}
\usetikzlibrary{decorations.markings}
\usepackage{algorithm}
\usepackage{algpseudocode}
\usepackage{mathrsfs}

{}
{}
{}

\begin{document}

\supertitle{}

\title{Minimum-Phase HRTF Modelling of Pinna Spectral Notches using Group Delay Decomposition}

\author{\au{Sandeep Reddy C$^{1\corr}$}, \au{Rajesh M Hegde$^{2}$}}
\address{\add{1,2}{Department of Electrical Engineering, Indian
Institute of Technology, Kanpur 208016, India}
\add{1}{Current affiliation: Center for Vision, Speech and Signal processing, University of Surrey}
\email{s.chitreddy@surrey.ac.uk, rhegde@iitk.ac.in}}

\begin{abstract}
Accurate reconstruction of HRTFs is important in the development of high quality binaural sound synthesis systems.
Conventionally, minimum phase HRTF model development for reconstruction of HRTFs has been limited to minimum phase-pure delay models which ignore the all pass component of the HRTF. In this paper, a novel method for minimum phase HRTF modelling of Pinna Spectral Notches (PSNs) using group delay decomposition is proposed.  The proposed model captures the PSNs contributed by both the minimum phase and all pass component of HRTF thus facilitating an accurate reconstruction of HRTFs. The purely minimum phase HRTF components and their corresponding spatial angles are first identified using Fourier Bessel Series method that ensures a continuous evolution of the PSNs. The minimum phase-pure delay model is used to reconstruct HRTF for these spatial angles. Subsequently, the spatial angles which require both the minimum phase and all pass components are modelled using an all-pass filter cascaded with minimum-phase pure-delay model. Performance of the proposed model is evaluated by conducting experiments on PSN extraction, cross coherence analysis, and binaural synthesis. Both objective and subjective evaluation results are used to indicate the significance of the proposed model in binaural sound synthesis. 
\end{abstract}

\maketitle

\section{Introduction}\label{sec1}

Sound interaction with human body varies with its direction of arrival. This interaction can be uniquely characterised by a direction dependent transfer function called as Head Related Transfer Function (HRTF) \cite{begault20003, blauert}. HRTFs are usually measured in an anechoic environment using a Loudspeaker-Microphone setup \cite{cipic,gardner}. These HRTFs measured for left and right ears can be utilised for synthesis of spatial sound.  Binaural synthesis is a headphone based spatial audio technique utilizing HRTFs of both left and right ear. In binaural synthesis, the HRTF is typically implemented as cascade of a minimum phase component of HRTF and a pure delay called as Minimum phase-pure delay model \cite{kulkarni,stanford}. This representation has been traditionally used since it reduces the cost of  binaural synthesis.  But it is well known that, spectral notches of HRTF, particularly contributed by pinna called as Pinna Spectral Notches (PSN), are considered to be dominant cues in the perception of sound directionality \cite{Wright}. Additionally, the contribution of these notches come from both minimum phase component and all pass component of HRTF. Therefore, an approximation of all pass component of HRTF using only a pure delay lead to loss in some of the spectral cues responsible for spatial sound perception. \footnote{\normalfont{This paper is a preprint of a paper submitted to IET Signal Processing Journal. If accepted, the copy of record will be available at the IET Digital Library}}

Spectral notches of HRTFs have been traditionally associated with the nulls of the HRTF magnitude spectrum. They also manifest as transitions in the phase spectrum. Phase transitions are also been contributed by the all pass component of HRTFs. All these transitions in the phase spectrum appear as sharp valleys in the group delay spectrum which is the negative differential of phase spectrum \cite{karan_MM}. In \cite{raykar}, a signal processing technique (LP-GD), employ autocorrelation, windowing, LP-residual and group delay operations on the HRTF to extract the location of the pinna spectral notches.  As group delay operates on the phase of HRTF, the notches observed in the group delay spectrum are due to minimum phase component and all pass component. It is important to understand the contributions from each component so that spectral notches which are lost in the minimum phase-pure delay model can be identified and captured by a compensation filter. However earlier perceptual studies indicate that the relevance of these subtle cues (spectral variations of all pass components) are perceptually indistinguishable when binaural synthesis is performed from a fixed direction \cite{kulkarni}. But it is important to study the relevance of the sudden variations of all pass phase in continuously rendering binaural audio. In order to address this problem, it is important to understand the phase variations continuously in the 3D space. Fourier Bessel Series can be utilised for studying these phase variations. Fourier Bessel Series were earlier used for interpolating horizontal plane HRTFs \cite{abhayahorz}. However spatial variation of all pass phase using FBS for any circular plane has not been explored in an explicit manner. In this study, we attempt to express HRTFs of any circular plane in Fourier Bessel Series. Subsequently HRTFs of any intermediate directions are reconstructed and their phase variations are studied using Group Delay decomposition. This helps us in understanding the evolution of spectral notches continuously and thereby identify the HRTFs and its corresponding directions which are purely minimum phase in nature.

The primary contributions of this work are as follows. 
A novel minimum phase HRTF model using group delay decomposition is proposed for binaural sound synthesis. In this context, a HRTF group delay decomposition method for  pure minimum phase HRTF identification using FBS method is first developed. Subsequently, a second order all pass filter is designed using specifications obtained by the phase variations of all pass component of HRTF. The HRTFs reconstructed by the proposed HRTF model with all pass filter are used for binaural sound synthesis using CIPIC, KEMAR, MiPS databases. Performance improvements are observed in terms of binaural sound perception.

The rest of the paper is organised in the following manner. Pure minimum phase HRTF identification using HRTF group delay decomposition is described in Section \ref{sec:2}. A FBS method for identification of pure minimum phase HRTFs is also discussed herein. Subsequently, HRTF reconstruction and spectral notch extraction for CIPIC database is also discussed in this section. Section \ref{sec:3} describes the proposed HRTF model using group delay decomposition and all pass filter for  binaural sound synthesis. Design of all pass filter is also discussed in this section. The performance of the proposed minimum phase HRTF model for PSN extraction and  binaural sound synthesis over CIPIC, KEMAR and MiPS databases is discussed in Section \ref{sec:4}. Section \ref{finalsec} concludes the paper.

\section{HRTF Group Delay Decomposition for Identification of Pure Minimum Phase HRTF}
\label{sec:2}

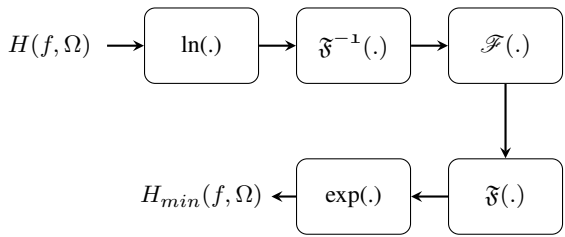
\begin{figure}[h]
%\centering{
\begin{tikzpicture}[node distance=2cm]
%\node (fig1) at (-2,0)
%       {\includegraphics[width=1.5cm,height=1.2cm]{recall11}};
%       \node (fig1) at (-2,-1.2)
%       {\includegraphics[width=1.5cm,height=1.1cm]{bb2}};
\tikzstyle{startstop} = [rectangle, rounded corners, minimum width=1.5cm, minimum height=1cm,text centered, draw=black, fill=white!30, align=center, below]
\tikzstyle{startstop1} = [rectangle, rounded corners,minimum width=1.5cm, minimum height=1cm,text centered, draw=white, fill=white!30, align=center, below]
\tikzstyle{arrow} = [thick,->,>=stealth]
\tikzstyle{arrow1} = [thick,<-,>=stealth]

\node (start1) [startstop1]  {$H(f,\Omega)$};
%\node[yshift=-1cm] at (current page.north west)
\node (start2) [startstop, right of=start1] {ln(.)};
\node (start3) [startstop, right of=start2] {$\mathfrak{F^{-1}(.)}$};
\node (start4) [startstop, right of=start3] { $\mathscr{F}(.)$};
\node (start5) [startstop, below of=start4] {$\mathfrak{F(.)}$};
\node (start6) [startstop, left of=start5] {exp(.)};
\node (start7) [startstop1, left of=start6] {$H_{min}(f,\Omega)$};
\draw [arrow]  (start1) -- (start2);
\draw [arrow]  (start2) -- (start3);
\draw [arrow]  (start3) -- (start4);
\draw [arrow]  (start4) -- (start5);
\draw [arrow1]  (start6) -- (start5);
\draw [arrow1]  (start7) -- (start6);
\end{tikzpicture}
\caption{Flow diagram for computing minimum phase component of HRTF}
\label{fig:minstep}
\end{figure}

Any transfer function can in general be decomposed into its minimum phase component and all pass component. Similarly, an HRTF measured from a direction $\Omega=(\phi,\theta)$ can also be represented as a cascade of its minimum phase component and pure delay component as, 
 
 \begin{equation}
H(f,\Omega)=H_{min}(f,\Omega) H_{ap}(f,\Omega)
\label{minapdecom}
\end{equation}
where $H_{min}(\cdot)$ is the minimum phase component of HRTF, and $H_{ap}(\cdot)$ is the all pass component of HRTF. $H(\cdot)$ is termed as composite HRTF herein. In principle, the minimum phase component of HRTF is obtained by factoring the transfer function polynomial and identifying zeros and poles which are inside unit circle. But factoring large polynomials is not practical and therefore a non parametrical method of obtaining minimum phase component is used herein \cite{FILTERS07,FILTERSWEB07}. The flow diagram for computing minimum phase component of HRTF is illustrated in Figure \ref{fig:minstep}. $\mathfrak{F}$ indicates the Fourier transform operator and $\mathscr{F}$ indicates the folding operation which is performed on complex cepstrum using Equations \ref{evencepstrum} and \ref{oddcepstrum} for even and odd N respectively.

\begin{equation}
\tilde{h}(k) = \begin{cases}
\hat{h}(k) &\text{ $k= 1,\frac{N}{2}+1$}\\
\hat{h}(k)+\hat{h}^{*}(N-k+2) &\text{ $k= 2 \cdots \frac{N}{2}$} \\
0 & \text{$k=\frac{N}{2}+2 \cdots N$} \\
\end{cases}
\label{evencepstrum}
\end{equation}

\begin{equation}
\tilde{h}(k) =
\begin{cases}
\hat{h}(k)  & \text{$k= 1 $} \\
\hat{h}(k)+\hat{h}^{*}(N-k+2)  & \text{ $k= 2 \cdots \frac{N+1}{2}$} \\
0 & \text{$k=\frac{N+3}{2} \cdots N$} \\
\end{cases}
\label{oddcepstrum}
\end{equation}

\begin{figure}[h]
\includegraphics[width=9cm,height=8.0cm]{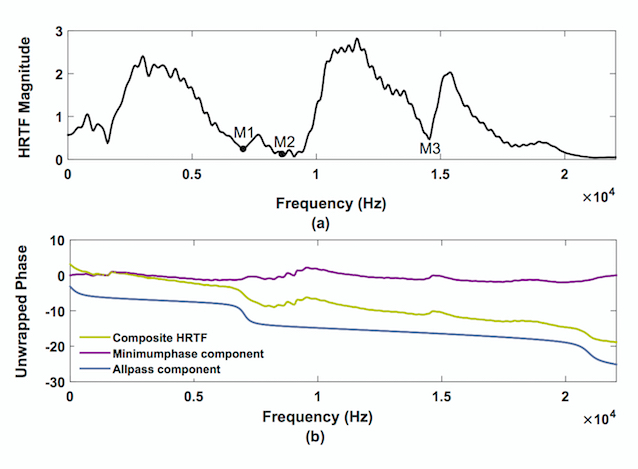}
\vspace{-1cm}
\caption{Magnitude and phase spectrum of subject 003 in the CIPIC database for angular direction $\Omega=(0^0,-11.25^0)$. (a) Magnitude of minimum phase HRTF and (b) Unwrapped phase spectrum of composite HRTF, minimum phase HRTF and all pass component of HRTF}
  \label{fig:Magandphase}
\end{figure}

\begin{figure}[t]
  \centerline{\includegraphics[width=10cm,height=8.5cm]{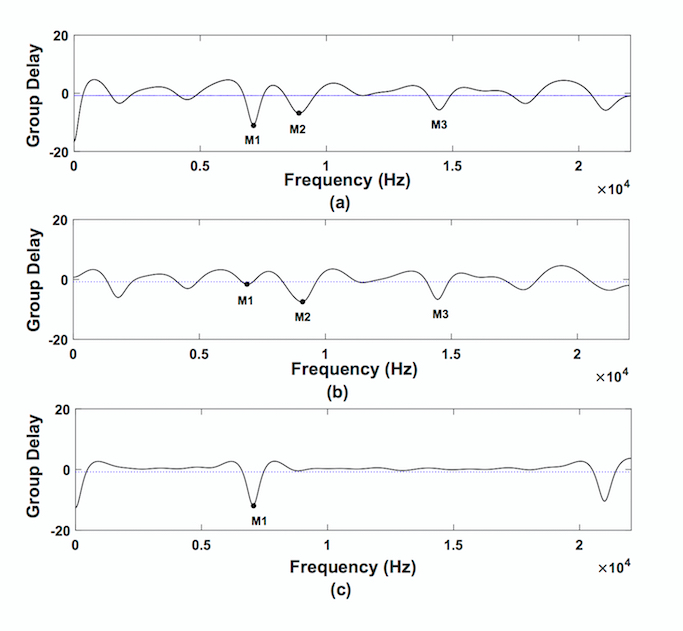}}
  %\vspace{-0.2cm}
  \caption{Figure illustrating spectral notches extracted using group delay decomposition. (a) Group delay spectrum of composite HRTF, (b) Group delay spectrum of minimum phase component of HRTF and (c) Group delay spectrum of all pass component of HRTF. HRTF corresponds to Subject 003 of CIPIC database for angular direction $\Omega=(0^0,-11.25^0)$}
  \label{fig:GD_sub3}
\end{figure}

A folding operation is performed to transform the non causal exponentials in the complex cepstrum to causal exponentials. In other words, exponentials due to the zeros and poles lying outside the unit circle are shifted to its conjugate reciprocals which lie inside the unit circle. After obtaining the minimum phase component, all pass component can be obtained as the ratio of composite HRTF and minimum-phase component of HRTF as in Equation \ref{minapdecom}. Fig. \ref{fig:Magandphase}(a) illustrates the magnitude spectrum of composite HRTF. Fig. \ref{fig:Magandphase}(b) illustrates the unwrapped phase spectrums of composite HRTF, minimum phase component of HRTF and all pass component of HRTF for Subject 003 of CIPIC database for angular direction $\Omega=(0^0,-11.25^0)$.  M1, M2 and M3 in Figure \ref{fig:Magandphase}(a) indicate the nulls in the magnitude spectrum. These nulls manifested as small  variations in the unwrapped phase spectrum of minimum phase component of HRTF which can be observed in Figure \ref{fig:Magandphase}(b). This relation stems from the fact that the magnitude response and phase of the minimum phase systems are related using a Hilbert operator \cite{FILTERS07,oppenheim2010discrete}. Further in this example, the all pass component also exhibit a phase variation. All these phase variations are manifested in the composite HRTF as the phase spectrum of composite HRTF is the sum of phase of minimum phase and all pass components. However some of the variations are indistinguishable in the phase spectrum.  In order to resolve this ambiguity, group delay decomposition of HRTFs is performed,
as the group delay spectrum exhibits sharper nulls when compared to the phase spectrum \cite{yegnanarayana1992significance, kumar2014robust}. 
In the rest of the section,  group delay decomposition of HRTF and pure minimum Phase HRTF identification is discussed.

\subsection{Decomposition of Group Delay of HRTF in to Minimum-phase and All Pass Components}
\label{GD-decomp}
The group delay function is defined as the negative derivative of phase spectrum. 
The additivity and high resolution property of group delay functions has been studied and applied widely in the domain of speech processing \cite{4032772}. These two properties of group delay functions can be used in the accurate  decomposition into minimum phase and all pass components.  The group delay of the composite HRTF can be decomposed in to its minimum-phase and all pass components as,

\begin{equation}
\tau_{g,com}=\tau_{g,min}+\tau_{g,ap}
\label{grdminap}
\end{equation}

\begin{figure*}[t]
  \centerline{\includegraphics[width=17cm,height=8.7cm,trim=30 60 0 40,clip]{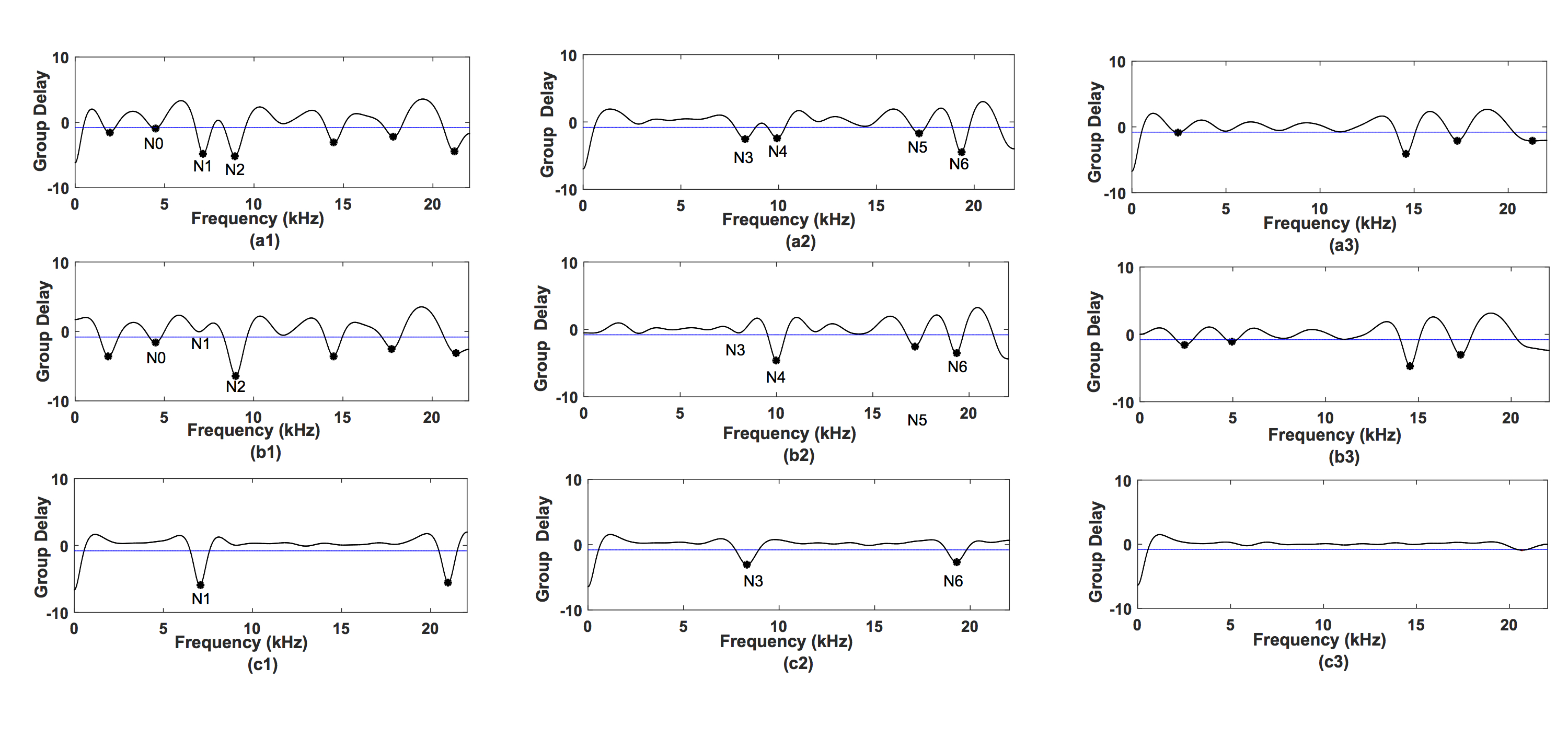}}
  %\vspace{-0.3cm}
\caption{Figures (a1,a2,a3) illustrate the group delay response of composite HRTF, Figures (b1,b2,b3) illustrates the group delay response of minimum-phase component of HRTF, Figures (c1,c2,c3) illustrates the group delay response of allpass component of HRTF for elevation angles $(-11.25^0, 16.875^0, 84.375^0)$ respectively corresponding to Subject 003 of CIPIC database.}
\label{fig:notch_classify}
\end{figure*}

where, $\tau_{g,com}$, $\tau_{g,min}$, and $\tau_{g,ap}$ are the group delay of composite, minimum phase and all pass components of HRTF respectively.
Group delay of each individual component can be obtained as \cite{oppenheim2010discrete},

\begin{equation}
\tau_g(\omega)=\frac{H_R(\omega)H^{'}_R(\omega)+H_I(\omega)H^{'}_I(\omega)}{H^{2}_R(\omega)+H^{2}_I(\omega)}
\end{equation}

where, $H_R(\omega)$ and $H_I(\omega)$ are the real and imaginary part of Fourier transform of $h[n]$ respectively. It may be noted that, $h_[n]$ is the Head Related Impulse Response (HRIR). $H^{'}_R(\omega)$ and $H^{'}_I(\omega)$ are the real and imaginary part of Fourier transform of $nh[n]$ respectively. The group delay response of composite, minimum phase and all pass components of HRTF are computed using a LP-GD based method as discussed in \cite{raykar}. The group delay response of all the three components is illustrated in Fig. \ref{fig:GD_sub3}(a)-(c) for Subject 003 of CIPIC database. Notches M2 and M3 in the group delay response of composite HRTF are clearly due to the nulls in the magnitude spectrum as shown in Fig. \ref{fig:Magandphase}(a). Notch M1 of group delay response of HRTF is contributed by the phase variation of all pass component that can be seen in Fig. \ref{fig:Magandphase}(b). The group delay magnitude of the composite HRTF at the notch frequencies can be attributed to the group delay additivity property given in Equation \ref{grdminap}. The significance of the group delay decomposition  in the identification of pure minimum phase HRTF is discussed in the ensuing section.

\subsection{Significance of Group Delay Decomposition in Identification of Pure Minimum-phase HRTF}
\label{GD-sig}
Group delay spectrum of composite HRTF is an addition of its minimum phase and all pass components as given in Equation \ref{grdminap}.  However, a composite HRTF is decomposed as the product of its minimum phase and all pass component as given in Equation \ref{minapdecom}. Hence, spectral notches can be resolved better in the group delay domain because of its additivity property \cite{kumar2014robust}. Additivity property of the group delay spectrum also helps in resolving the closely spaced spectral notches of the composite HRTFs. In order to illustrate the significance of the group delay spectrum, consider HRTFs for elevation angles $(-11.25^0, 16.875^0, 84.375^0)$ respectively of Subject 003 of CIPIC database. The group delay spectrum of the composite HRTF and, its minimum phase and all pass components, for all the three elevation angles are shown in Fig. \ref{fig:notch_classify}. Three different cases are illustrated in Fig. \ref{fig:notch_classify}. In the first case shown in Fig. \ref{fig:notch_classify} (a3)-(c3) that, spectral notches are contributed only by the minimum phase component and not by the all pass component.  The second case can be categorised based on the location of spectral notches in the minimum phase and all pass components.  It can be observed from Fig. \ref{fig:notch_classify} that, notches N1 and N3 in Fig. \ref{fig:notch_classify}(a1)  and \ref{fig:notch_classify}(a2) respectively indicate the spectral notches due to the all pass component of HRTF. Notches N2, N4 and N5 indicate the spectral notches contributed by the minimum phase component of HRTF.  HRTFs can therefore be classified into Pure Minimum-phase HRTFs and Minimum phase-All pass HRTFs

\subsubsection{Pure Minimum-phase HRTFs}
HRTFs of  spatial angles which are completely described by spectral notches due to the minimum phase component and not by the all pass component are termed as Pure-Minimum phase HRTFs. HRTFs of these directions can be modelled as a cascade of minimum phase component and a pure delay component. The resultant minimum phase HRTF preserves all spectral notches of the composite HRTF. HRTF exhibiting pure minimum phase behaviour is shown in Fig. \ref{fig:notch_classify} (a3)-(c3). 

\subsubsection{Minimum phase-All pass HRTFs}

HRTFs of spatial angles which are completely described by spectral notches by both the minimum phase and all pass components of HRTF are called Minimum phase-All pass HRTFs. A minimum phase-All pass HRTF is shown in Figs. \ref{fig:notch_classify}(a1)-(c1) and (a2)-(c2). HRTFs of these spatial directions cannot be modelled as a cascade of minimum phase component and pure delay component as the spectral notches of the all pass component cannot be captured by the minimum phase-pure delay model. Therefore it is important to develop a minimum phase HRTF model which can capture the spectral notches of the all pass component. Before introducing this model, identification of pure minimum HRTFs and its corresponding spatial angles using FBS method is discussed.

\subsection{Indentification of Pure Minimum-phase HRTFs and Corresponding Spatial Angles using FBS Method}
\label{TLFBS}
Identification of pure minimum-phase HRTFs and its corresponding spatial angles is an important step in the development of a new minimum phase HRTF model. Identification of pure minimum phase HRTFs of various spatial angles is performed herein, by analytical modelling of composite HRTFs.  Fourier Bessel Series (FBS) is one method to model composite HRTFs of a circular plane. FBS method was earlier studied in the horizontal
plane HRTF interpolation \cite{abhayahorz}. However FBS method is used here to reconstruct HRTFs of any arbitrary spatial direction in a particular circular plane. Spectral notches are extracted for the reconstructed HRTFs, and the range of azimuthal and elevation angles for which all pass spectral notches exist are identified. The following sections details the FBS based HRTF modelling and identification of purely minimum phase HRTFs. 

\subsubsection{FBS based Modelling of Composite HRTF}
\label{FBS_Modelling}

Consider HRTFs of equally spaced discrete directions on any circle circumscribed over a sphere. Fourier series representation of these HRTFs is given by

%\vspace{-0.25cm}
\begin{equation}
 H(f,\theta) = \sum\limits_{m=-\infty}^{\infty} C_{m}(f)e^{jm\theta} \quad  0\leq\theta<2\pi
 \label{Fseries}
\end{equation}
%\vspace{-0.25cm}

Fourier series coefficients in Equation \ref{Fseries} are frequency dependent. These coefficients capture the resonances and notches present in HRTF. It has been shown in earlier work that spectral component of HRTF have similarities with the Bessel functions \cite{abhaya}. Utilising this knowledge,  $C_{m}(f)$ is represented using Bessel functions as,

%\vspace{-0.5cm}

\begin{align}
\label{eq:6}
C_{m}(f)=\sum\limits_{k=1}^{\infty} C_{mk}J_{|m|}(\beta_{k}^{|m|}\frac{f}{f_{max}})
\end{align}

where, $J_{n}(.)$ is the Bessel function of first kind and order n. $\beta_{1}^{n},\beta_{2}^{n}\cdots, \beta_{k}^{n}$ are the positive roots of $J_{n}(x)=0$. Substituting Equation \ref{eq:6} in to Equation \ref{Fseries} gives a combined orthogonal representation of  HRTFs as,
\begin{align}
\label{eq:7}
H(f,\theta) = \sum\limits_{m=-\infty}^{\infty}\sum\limits_{k=1}^{\infty} C_{mk}J_{|m|}(\beta_{k}^{|m|}\frac{f}{f_{max}})e^{jm\theta}
\end{align}
The complex coefficients $C_{mk}$ are found by utilising the orthogonal property of Bessel functions. In Equation \ref{eq:7}, for various values of k, Bessel functions are orthogonal functions. The orthogonal property of Bessel functions is given below.
\begin{align}
\label{eq:8}
\int_{x=0}^{1}xJ_{|l|}(\beta_{k}^{|l|}x)J_{|l|}(\beta_{k'}^{|l|}x)dx = \frac{[J_{|l|+1}(\beta_{k}^{|l|})]^{2}}{2}\delta_{kk'}
\end{align}

Further for various values of m, the exponential terms are orthogonal. Therefore applying joint orthogonality property, coefficients $C_{mk}$ can be found as,

\begin{align}
\label{eq:9}
C_{mk} = \frac{1}{\pi[J_{|m|+1}(\beta_{k}^{|m|})]^{2}}\int\limits_{0}^{f_{max}}\int\limits_{-\pi}^{\pi} \frac{fH(f,\theta)}{f_{max}^{2}} &J_{|m|}(\beta_{k}^{|m|}\frac{f}{f_{max}}) \nonumber\\ \cdots e^{-jm\theta}dfd\theta
\end{align}

\begin{figure}
\begin{center}
   \centering
   \centerline{\includegraphics[width=9cm,height=7.0cm,trim=0 0 0 30,clip]{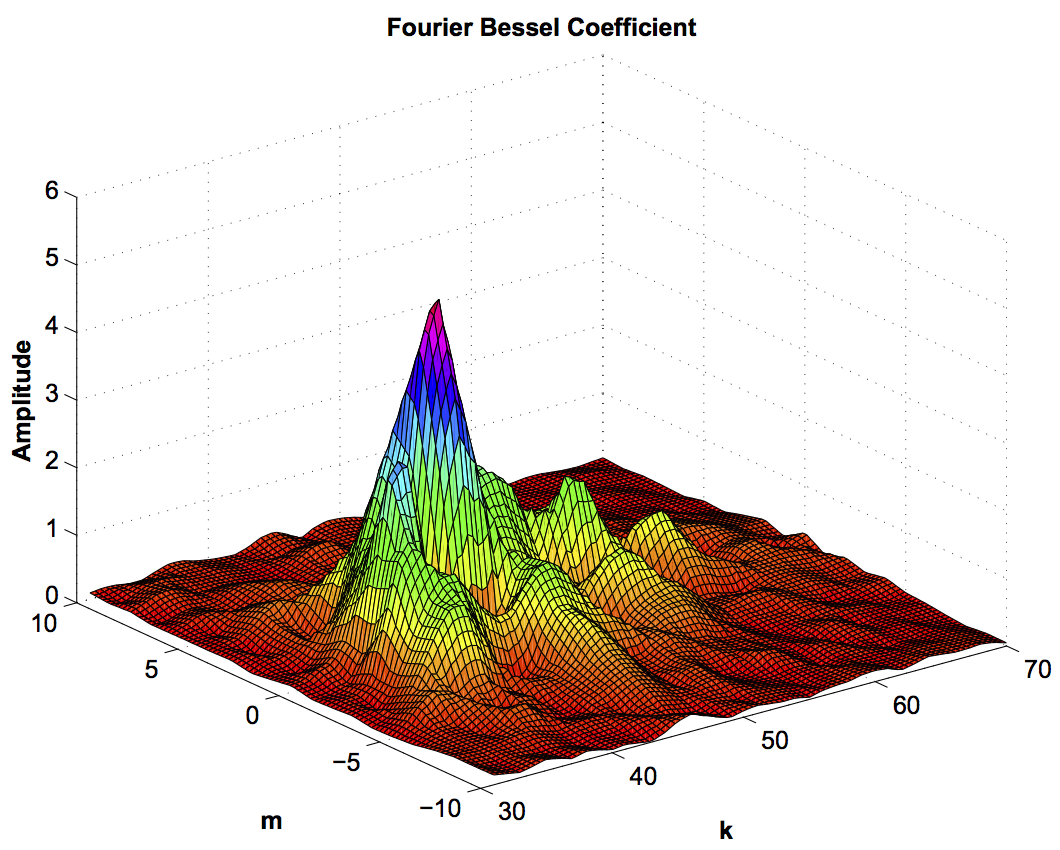}}
   \vspace{0.45cm}
   \caption{Amplitude of Fourier Bessel coefficient $|C_{mk}|$ in the median plane for left pinna of subject 50 in CIPIC Database.}
   \label{fig:fourier_bessel_coeff}         
\end{center}
%\hfill
\end{figure}

Equation \ref{eq:9} consists of continuous functions and an integration operator.
In practice the integral needs to be computed by discrete approximation of Equation \ref{eq:9}. A discrete approximation of Equation \ref{eq:9} can be obtained by

\begin{align}
\label{eq:11}
C_{mk} = \frac{1}{MN[J_{|m|+1}(\beta_{k}^{|m|})]^{2}f_{max}}\sum\limits_{f_{i}=0}^{f_{max}}\sum\limits_{\phi_{i}=-{\pi}}^{{\pi}} &f_{i}H(f_{i},\theta_{i}) \cdots \nonumber\\  \cdots J_{|m|}(\beta_{k}^{|m|}\frac{f_{i}}{f_{max}})e^{-jm\theta_{i}}
\end{align}

where M and N denote the total number of discretised frequency bins and angular directions respectively. Computation of $C_{mk}$ for all $m$ and $k$ is practically not possible.  Therefore truncation to a  finite value is required. This truncation is justified because the magnitude of $C_{mk}$ coefficients is significant for only finite $m$ and $k$ values as illustrated in Fig. \ref{fig:fourier_bessel_coeff}. It can be inferred from the Fig. \ref{fig:fourier_bessel_coeff} that, for $\left| m \right|>10, k>70$ and $k<30$, $C_{mk}$ magnitude is considerably very less.  Using the truncation limits HRTFs can be reconstructed for any arbitrary elevation angles. In this manner HRTFs thus computed from FBS method can further be used to obtain the spectral notches for any arbitrary angle $(\theta,\phi)$. In this work reconstruction is performed for all the elevation angles of all the 25 interaural circles of CIPIC database. 

\begin{figure}
\begin{center}
   \centering
   \hspace{0.8cm}\centerline{\includegraphics[width=9.5cm, height=15cm]{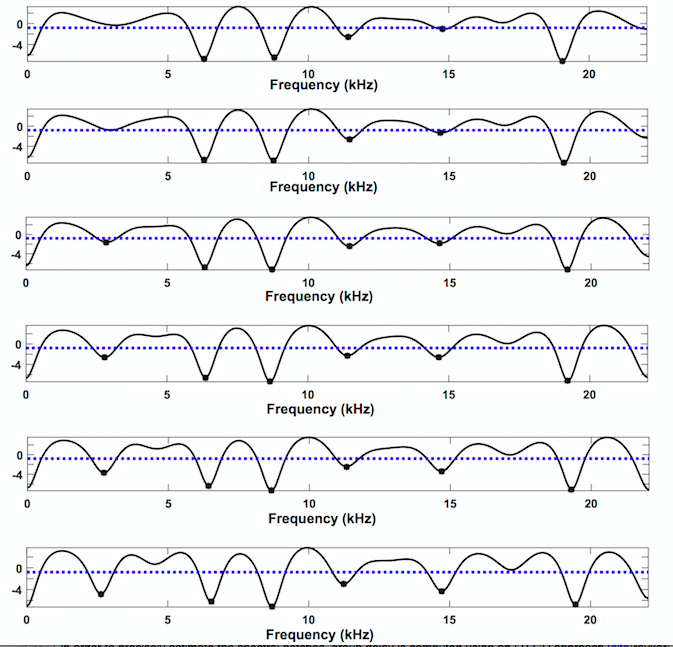}}
   %\vspace{-0.6cm}
   \caption{Group delay response of reconstructed HRTF for elevation angles ranging from $-39.37^o$ to $-33.75^o$ in steps of $1.125^0$ (top to bottom). Spectral notches are indicated using bold dots.}
   \label{fig:GD}         
\end{center}
%\hfill
\end{figure}

\begin{figure}
\centerline{\includegraphics[width=10.5cm,height=6cm]{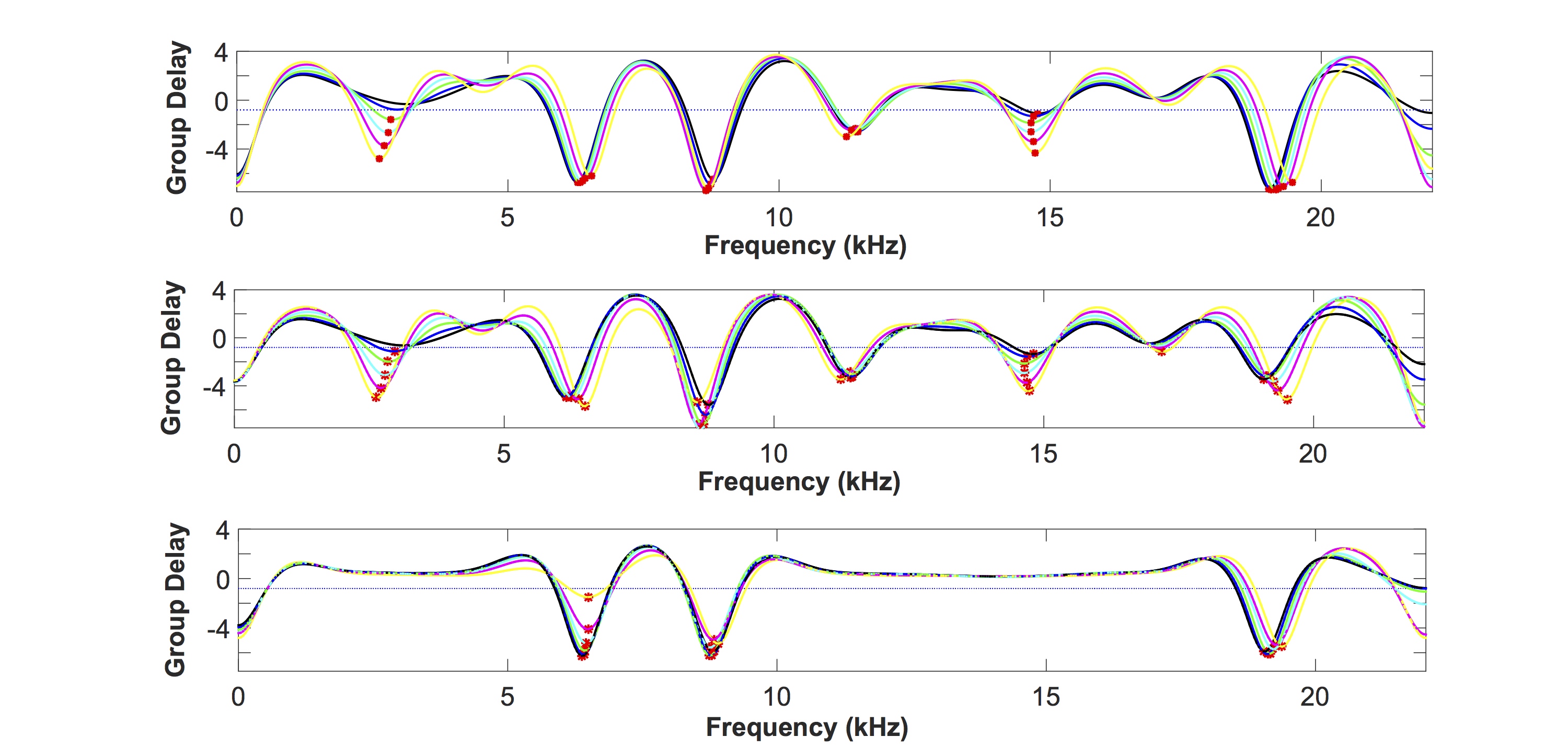}}
%\vspace{-0.3cm}
\caption{ PSNs extracted for  reconstructed HRTFs, and their minimum-phase and allpass components for elevation angles ranging from $-39.37^0$ to $-33.75^0$. Elevation angle step size of $1.125^0$ is used.}
\label{fig:FBS_minexcess}
\end{figure}

In order to precisely estimate the spectral notches, group delay is computed using an LP-GD approach \cite{raykar}. This algorithm is used to obtain the location of spectral notches, and notch depth.  In order to distinguish notches a threshold of -0.8dB is used in this work. Fig. \ref{fig:GD} illustrates the group delay plot obtained for the HRTFs reconstructed for an elevation angle $-39.37^{\circ}$ to $-33.75^{\circ}$ in steps of $1.125^{\circ}$. It can observed clearly that there is a monotonic increase of the depth of notch 1 and notch 5. It may be noted that, the HRTF reconstructed using FBS for any arbitrary elevation angle has a continuous evolution in terms of spectral notches with respect to its preceding and succeeding elevation angles.

\begin{figure*}[t!]
  \centerline{\includegraphics[width=20cm,height=10.5cm]{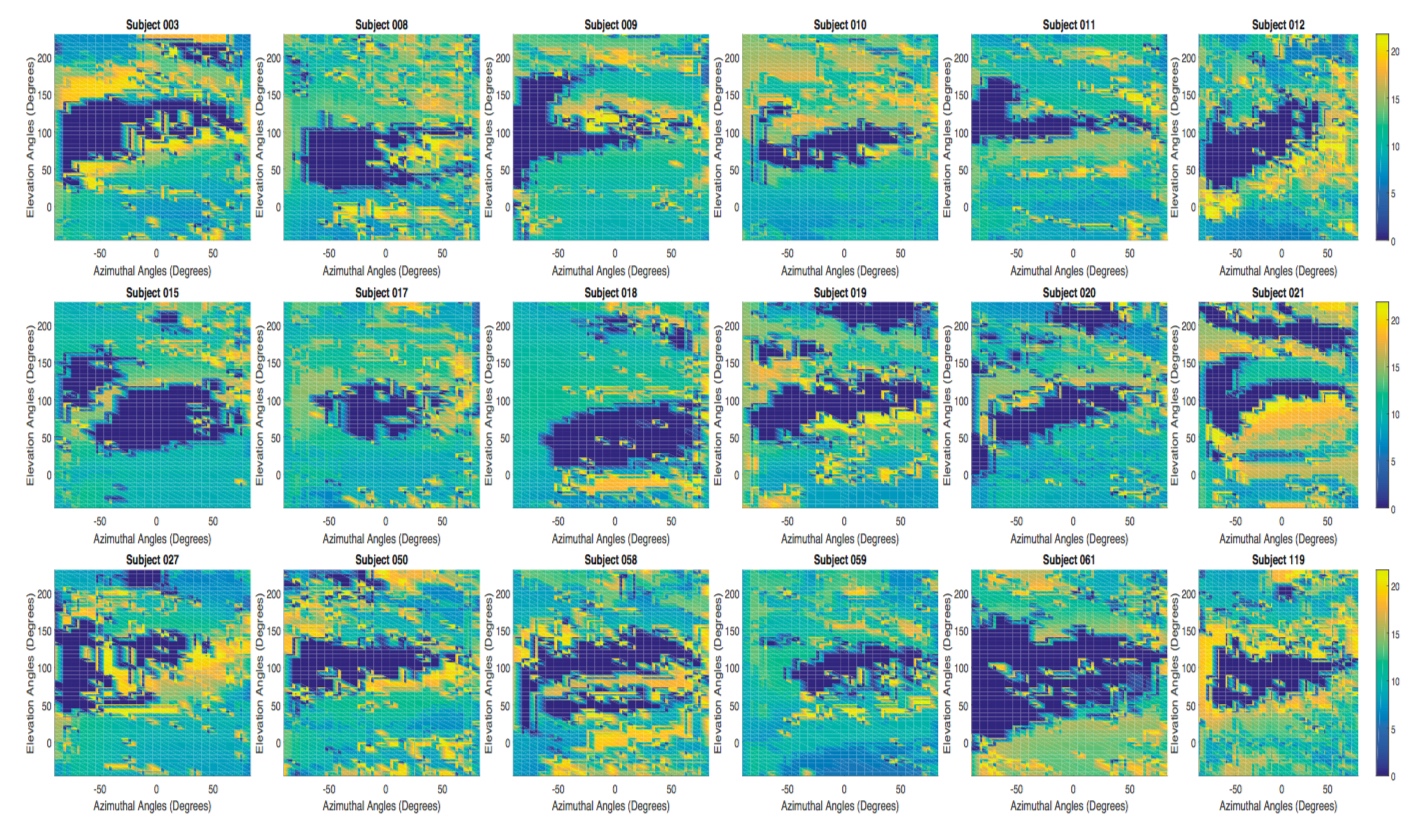}}
\caption{Location of the all pass spectral notch in the audio frequency band from 20Hz to 20KHz extracted for the HRTFs of various subjects of CIPIC database.}
\label{fig:allpi2}
\end{figure*}

\subsubsection{Identification of Pure Minimum phase HRTFs from Composite HRTFs}

FBS based modelling of composite HRTFs can be used to reconstruct composite HRTFs of any elevation angle in an interaural circle. This result of the FBS method indicate that it is possible to observe the spatial dynamics of the notches continuously in the 3D space. Hence it is also possible to observe the spatial dynamics of minimum phase and all pass spectral notches so that purely minimum phase HRTFs can be identified. An example illustrating the spatial dynamics of minimum phase and all pass spectral notches is given in Fig. \ref{fig:FBS_minexcess}. It can be understood from the first notch of all pass component that, FBS method can exactly identify the angular direction at which all pass spectral notches crosses the threshold.  The steps  for finding the pure minimum phase HRTFs and corresponding spatial angles is enumerated in Algorithm \ref{alg1}. The inputs to Algorithm \ref{alg1} are the set of measured HRIRs. The algorithm utilizes the FBS method described in Section \ref{FBS_Modelling} to output $\Omega$ and $\mathscr{H}$. $\Omega$ is the range of all angles for which HRTFs exhibit purely minimum phase behaviour in a particular circular plane, and $\mathscr{H}$ includes corresponding  minimum-phase HRTFs. This algorithm can be repeated for all the azimuthal angles to identify the pure minimum phase HRTFs of all directions of 3D space.

Simulations are performed on HRTFs of CIPIC database across various subjects using proposed algorithm to identify the purely minimum phase HRTFs. Figure \ref{fig:allpi2} indicates the depth of the all pass spectral notch for various directions. The color (Blue) in Figure \ref{fig:allpi2} indicates the regions where HRTFs does not exhibit any spectral notch by the all pass component and therefore considered as purely minimum phase in nature. It can be observed that HRTFs corresponding to roughly azimuthal angles ranging from $-90^0$ to $90^0$ and elevation angles ranging from $50^0$ to $120^0$, exhibit pure minimum phase behaviour.

\begin{algorithm}[h]
\caption{Algorithm to identify pure minimum phase HRTFs and its corresponding directions}
\label{alg1}
\begin{algorithmic}[1]
\State {\bf Input:} $\theta_0= 0^0$, $\phi$, Set of measured HRIRs $\mathscr{D}$
\State HRTF  reconstruction using  FBS $\rightarrow$ $H(f,\theta_0,\phi)$
\State Extract minimum phase and all pass components $H_{min}(f,\theta_0,\phi)$ and $H_{ap}(f,\theta_0,\phi)$
\State Compute the group delay of all pass HRTFs, say $\tau_{ap}$.
\If {$min(\tau_g)<-0.8$}
    \State $\Omega=\Omega \cup (\theta_0,\phi)$
    \State $\mathscr{H}=\mathscr{H}\cup \{H(f,\theta_0,\phi)\}$
    \If {$\theta_0\not=360^0$}
    \State $\theta_0\gets \theta_0+1^0$
    \State Jump to Step 2
%    \ElsIf {$\theta_0=180^0, \phi_0\not=360^0$}
%    \State $\phi_0\gets \phi_0+1^0, \theta_0=0^0$
%    \State Jump to step 2
    \Else
    \State STOP
    \EndIf
\EndIf
\State {\bf Output:} Set of purely minimum phase HRTFs $\mathscr{H}$, and its corresponding directions $\Omega$.
\end{algorithmic}
\end{algorithm}

\section{A HRTF Model using Minimum-phase Decomposition for  Binaural Sound Synthesis}
\label{sec:3}
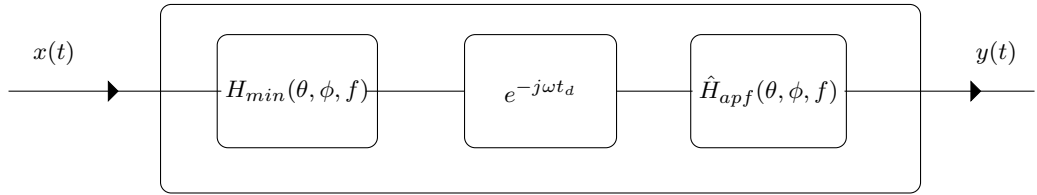
\begin{figure*}[t!]
\newcommand{\midarrow}{\tikz \draw[-triangle 90](0,0) -- +(.1,0);}
\begin{tikzpicture}[node distance=2.4cm]
\hspace{-1cm}
\tikzstyle{arrow} = [thick,->,>=stealth]
\tikzstyle{arrow1} = [thick,<-,>=stealth]
\tikzstyle{arrow2} = [thick,-,>=stealth]
\tikzstyle{arrow3} = [dashed,-,>=stealth]
 % \node [draw,rectangle, rounded corners, minimum width=10cm,minimum height=3.5cm] at (12,1.6){}; 
 \node [draw,rectangle, rounded corners, minimum width=10cm,minimum height=2.5cm] at (12,-2.6){}; 
  %\node [draw,rectangle, rounded corners, minimum width=12cm,minimum height=9.5cm] at (12,-0.4){}; 

%  \node [draw,rectangle, rounded corners, minimum width=2cm,minimum height=1.5cm] at (9.5,1.6){$H_{min}(\theta,\phi,f)$}; 
% \node [draw,rectangle, rounded corners, minimum width=2cm,minimum height=1.5cm] at (14,1.6){$e^{-j\omega t_d}$};
 
%   \node at (12,0.2) (text)  {$(\theta,\phi) \in \Omega$ \quad \quad MP Model};
%   \node at (12,-4) (text)  {$(\theta,\phi) \notin \Omega$ \quad \quad MP-MNF Model};

\node at (5.6,-2) {$x(t)$}; 
   
   \node [draw,rectangle, rounded corners, minimum width=2cm,minimum height=1.5cm] at (8.8,-2.5){$H_{min}(\theta,\phi,f)$}; 
     \node [draw,rectangle, rounded corners, minimum width=2cm,minimum height=1.5cm] at (12,-2.5){$e^{-j\omega t_d}$}; 
 \node [draw,rectangle, rounded corners, minimum width=2cm,minimum height=1.5cm] at (15,-2.5){$\hat{H}_{apf}(\theta,\phi,f)$};

%\draw [arrow3]  (5,-0.5) -- (19,-0.5);

%\begin{scope}[very thick, every node/.style={sloped,allow upside down}]
%  \draw (3,1.5)-- node {\midarrow} (7.7,1.5);
%   \draw (7.7,1.5)-- (8.4,1.5);
%  \draw (10.65,1.5)-- (13,1.5);
%  \draw (15,1.5)-- (17.7,1.5);
%    \draw (17.7,1.5)-- node {\midarrow} (19.5,1.5);
  
  \draw (2.5,-2.5)--(2.5,-2.5);
   \draw (5,-2.5)-- node {\midarrow} (7.8,-2.5);
  \draw (9.7,-2.5)-- (11,-2.5);
  \draw (13,-2.5)--(14,-2.5);
  \draw (16,-2.5)-- (17,-2.5);
    \draw (17,-2.5)-- node {\midarrow} (18.5,-2.5);
    \node at (18,-2) {$y(t)$}; 
    
%\end{scope}

 %\node at (12,-4.3) {(a)};

\end{tikzpicture}

%\vspace{-0.5cm}
\caption{ Figure illustrates cascading of minimum phase component of HRTF, pure delay, and second order all pass filter.}
\label{fig:min-pdflow}
\end{figure*}

  In spatial directions where HRTFs do not exhibit pure minimum phase behaviour, there is a contribution of spectral notches from the all pass component. Therefore a new minimum phase HRTF model is studied  so that the resultant HRTF captures all spectral cues. The minimum phase HRTF model is illustrated in Fig. \ref{fig:min-pdflow}. The model consists of three components that are cascaded in series. The first component, $H_{min}(\theta,\phi,f)$ indicates the minimum phase component of composite HRTF. The second component $e^{-j\omega t_d}$ indicates the pure delay, where $t_d$ indicates the onset time of the measured HRIRs.  The third component $\hat{H}_{apf}(\theta,\phi)$ is the second order all pass filter that models all pass component $H_{ap}(\theta,\phi)$. It must be noted that, for $(\theta,\phi) \in \Omega$, the all pass components do not exhibit spectral notches and hence the transfer function of APF has no zeros and poles. But for directions where $(\theta,\phi) \notin \Omega$, the transfer function contains zeros and poles. Hence the resultant transfer function of the proposed HRTF model $H_r(\theta,\phi)$ is given by

\[
H_{r}(\theta,\phi)=
\begin{cases}
H_{min}(\theta,\phi) e^{-j\omega t_d} &  \text{if } (\theta,\phi) \in \Omega \\ 
H_{min}(\theta,\phi) e^{-j\omega t_d}  \hat{H}_{apf}(\theta,\phi) & \text{Otherwise} \\
\end{cases}
\]

The design of the all pass filter $\hat{H}_{apf}(\theta,\phi)$ is discussed in the ensuing section.

\begin{figure*}[b!]
  \centerline{\includegraphics[width=20cm,height=10.0cm]{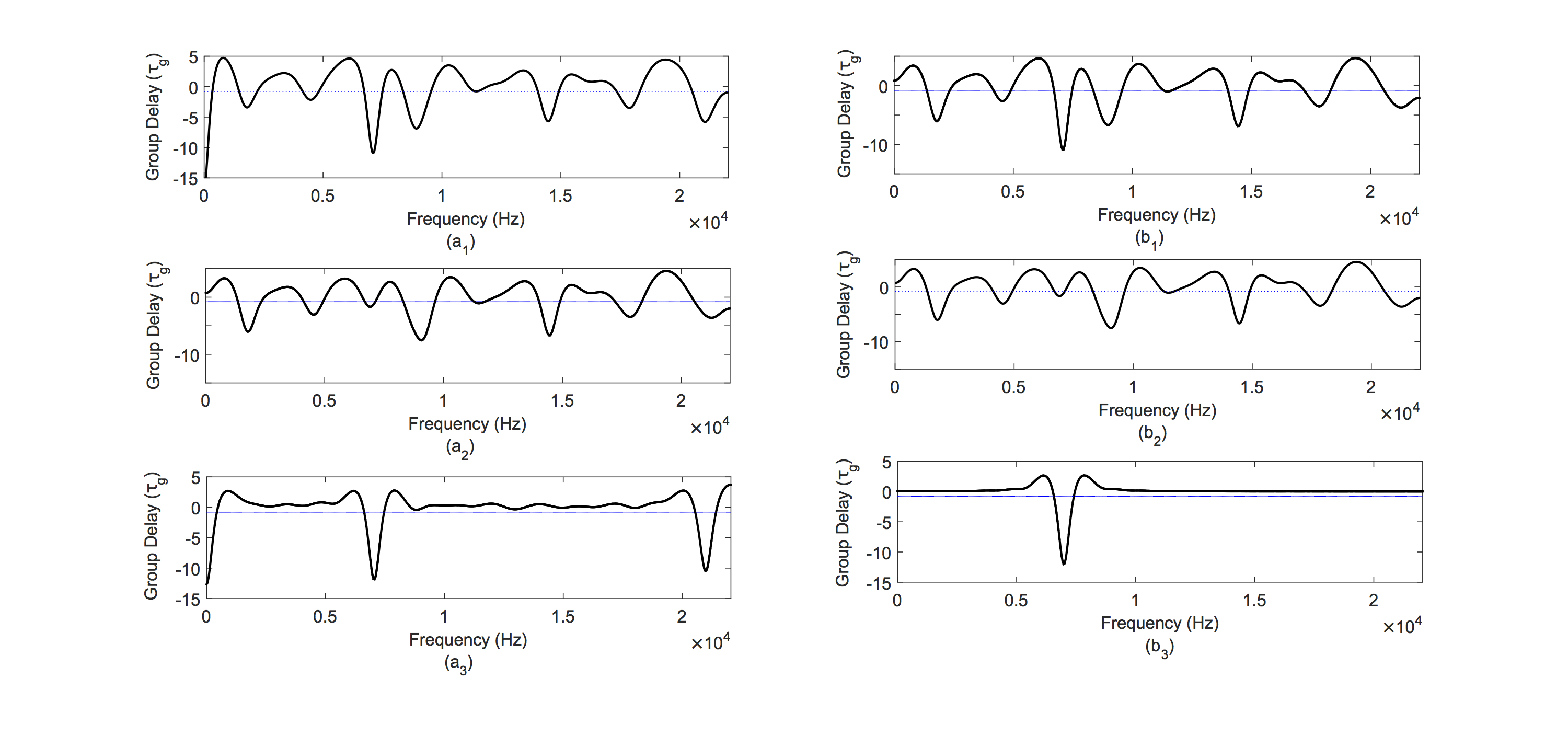}}
  \vspace{-0.8cm}
\caption{Figure $(a_1, a_2, a_3)$ illustrating the variation of group delay magnitude of composite HRTF, minimum phase component and all pass component of HRTF respectively, and Figure $(b_1, b_2, b_3)$ illustrates composite HRTF constructed using second order all pass filter $(r = 0.96, f_0 = 6991Hz)$, minimum phase component of $(b_1)$, and second order all pass filter respectively obtained using LP-GD approach.}
\label{fig:GD_apvsapf}
\end{figure*}

\subsection{Design of a Second Order All Pass Filter for Modelling All Pass Component of HRTF}
All pass component of any transfer function exhibits a flat magnitude spectrum. The differences in the all pass component of various transfer functions can  only be observed in the phase response. In general, a second order all pass transfer function can be expressed as,

\begin{eqnarray}
\label{notchsecond1}
\hat{H}_{apf}(z)=\frac{(z^{-1}-c^*_1)}{(1-c_1 z^{-1})}\frac{(z^{-1}-c^*_2)}{(1-c_2 z^{-1})}  \\
  c_1=re^{j\theta_0}, c_2=re^{-j\theta_0} \quad 
 \theta_0=\frac{2\pi f_0}{f_s} \label{phaserel} 
\end{eqnarray}

where $c_1,\frac{1}{c^*_1}$ are conjugate reciprocal to each other and they correspond to locations of pole and zero respectively. $(r,\theta_0)$ are their corresponding magnitudes and angles. $f_0$ represent frequency location where the all pass filter exhibit a phase variation, and $f_s$ corresponds to sampling frequency. This transfer function exhibit flat magnitude spectrum, however phase and its corresponding group delay of the transfer function depends upon the magnitude and angle of the $c_1, c_2$. The angle $\theta_0$ is determined as given in Equation \ref{phaserel}. Magnitude $r$ is determined by  the depth of notch that is observed in the group delay of all pass component. The phase and group delay spectrum of an all pass filter is related to its magnitude of the pole $r$ as follows.\begin{eqnarray}
\angle{\hat{H}_{apf}(\omega)}=\Phi=-2\omega-2\tan^{-1}\bigg(\frac{r\sin(\omega-\theta_0)}{1-r \cos(\omega-\theta_0)}\bigg) \nonumber \\ -2\tan^{-1}\bigg(\frac{r\sin(\omega+\theta_0)}{1-r \cos(\omega+\theta_0)}\bigg) 
\label{notchsecond2}
\end{eqnarray}

%\vspace{-0.5cm}

\begin{eqnarray}
\label{notchsecond3}
\tau_{g} = \frac{1-r^2}{1+r^2-2r\cos(\omega+\theta_0)}+\frac{1-r^2}{1+r^2-2r\cos(\omega-\theta_0)}
\end{eqnarray}
%\vspace{-0.8cm}

\begin{eqnarray}
\label{notchsecond4}
\omega=\pm \theta_0, \quad \tau_{g}=-\frac{d\Phi}{d \omega}  = \frac{1+r}{1-r}+\frac{1-r^2}{1+r^2-2r\cos(2\theta_0)}
\end{eqnarray}
It can be observed that the relation between the depth of the notch and the magnitude of $r$ is not linear. Location of the notches obtained using LP-GD approach are used for finding the angle $\theta_0$ and $\tau_g$. For a fixed $\tau_g$, and $\theta_0$, the polynomial expression in $r$ can be solved to obtain the magnitude of pole $r$. In this manner second order all pole filter is designed. 

Experiments are performed using HRTFs of Subject 003 of CIPIC database for angular direction $\Omega=(0^0,-11.25^0)$. Figure \ref{fig:GD_apvsapf} $(a_1, a_2, a_3)$ illustrates the group delay of the composite HRTF, minimum phase component of HRTF, and all pass component of HRTF obtained using LP-GD approach. It can be noted that, the all pass component of HRTF consists of a spectral notch at frequency $f_0=6991 Hz$. As group delays of minimum phase and all pass components are additive, the spectral notch contributed by the all pass component is manifested at the same location in the composite HRTF. The HRTFs generated using the minimum phase-pure delay model does not capture the contribution of these notches that can be noted from Figure $a_2$. Figure \ref{fig:GD_apvsapf} $(b_3)$ indicates the group delay of second order all pass filter using $(r = 0.96, f_0 = 6991Hz)$ obtained using LP-GD approach. Figure \ref{fig:GD_apvsapf} $(b_2)$ indicate the group delay of the minimum phase component of HRTF.  Figure \ref{fig:GD_apvsapf} $(b_1)$ indicates the group delay of the final composite HRTF obtained using the proposed method. It can be noted that, all spectral notches of minimum phase or all pass components are well captured.

\subsection{Significance of minimum phase HRTFs in  binaural sound synthesis}
Binaural synthesis is usually performed by filtering any monophonic sound with Left and Right HRTFs. It is important to understand the advantages of using minimum phase HRTFs as compared to composite HRTFs in the  binaural sound synthesis. Minimum phase systems reduces the cost of binaural synthesis by shortening the length of FIR filters and reducing the components required in the linear decomposition methods. Hence for the directions whose HRTFs exhibit purely minimum phase nature, the cost of binaural synthesis will be less as they can be completely represented in minimum phase components. For other directions a second order all pass filter adds additional cost but comes at an advantage of capturing important spectral cues. So instead of approximating all the HRTFs using minimum phase component and pure delay, in this work we selectively identify the directions for minimum phase representation thereby reducing the complexity and also capturing the important cues that are relevant for spatial sound perception. Considering these advantages the HRTFs obtained using the proposed method are used in the  binaural sound synthesis. The steps involved in obtaining the proposed HRTFs and the  binaural sound synthesised using these HRTFs are enumerated in Algorithm \ref{alg2}. The ensuing section discusses the performance of the proposed HRTFs as compared to HRTFs obtained using minimum phase pure delay model.

\begin{algorithm}[h]
\caption{Algorithm for  Binaural Sound Synthesis using the proposed HRTF model}
\label{alg2}
\begin{algorithmic}[1]
\State {\bf Input:}  $\theta_0$, $\phi_0$, purely minimum-phase HRTFs $(\mathscr{H})$, and its corresponding spatial angles $(\Omega)$, and monophonic sound $x(t)$.
\State Extract minimum phase and all pass components $H_{min}(f,\theta_0,\phi_0)$ and $H_{ap}(f,\theta_0,\phi_0)$
\State Compute the time delay $t_d$ using the onset time of HRIR.
\State  Compute the zeros $\mathbf{z}^*=\{z_1, z_2\}$, and poles $\mathbf{p}^*=\{p_1,p_{2}\}$  of second order all pass filter using all pass spectral notch specifications.
\State Using the zeros and poles obtain the second order all pass transfer function $\hat{H}_{apf}(\theta_0,\phi_0)$ 
\If {$(\theta_0$, $\phi_0) \in \Omega$}
\State $H_{r}(\theta_0,\phi_0)=H_{min}(\theta_0,\phi_0) e^{-j\omega t_d}$
\Else
\State $H_{r}(\theta_0,\phi_0)=H_{min}(\theta_0,\phi_0) e^{-j\omega t_d} \hat{H}_{apf}(\theta_0,\phi_0)$
\EndIf
\State Compute the left and right minimum phase HRTFs $H_{r,l}$ and $H_{r,r}$ and its corresponding HRIRs $h_{r,l}$ and $h_{r,r}$ using Steps 2 to 10.
\State  $y_l(t)=x(t)*h_{r,l}$, \quad $y_r(t)=x(t)*h_{r,r}$
\State Play the left and right channel output through headphones.
\State {\bf Output:} HRTFs obtained using the proposed method $H_{r,l}$, and $H_{r,r}$, and Binaural sound.
\end{algorithmic}
\end{algorithm}

\begin{figure*}
\begin{center}
   \centering
\centerline{\includegraphics[width=18cm, height=11.5cm,trim=100 10 10 10,clip]{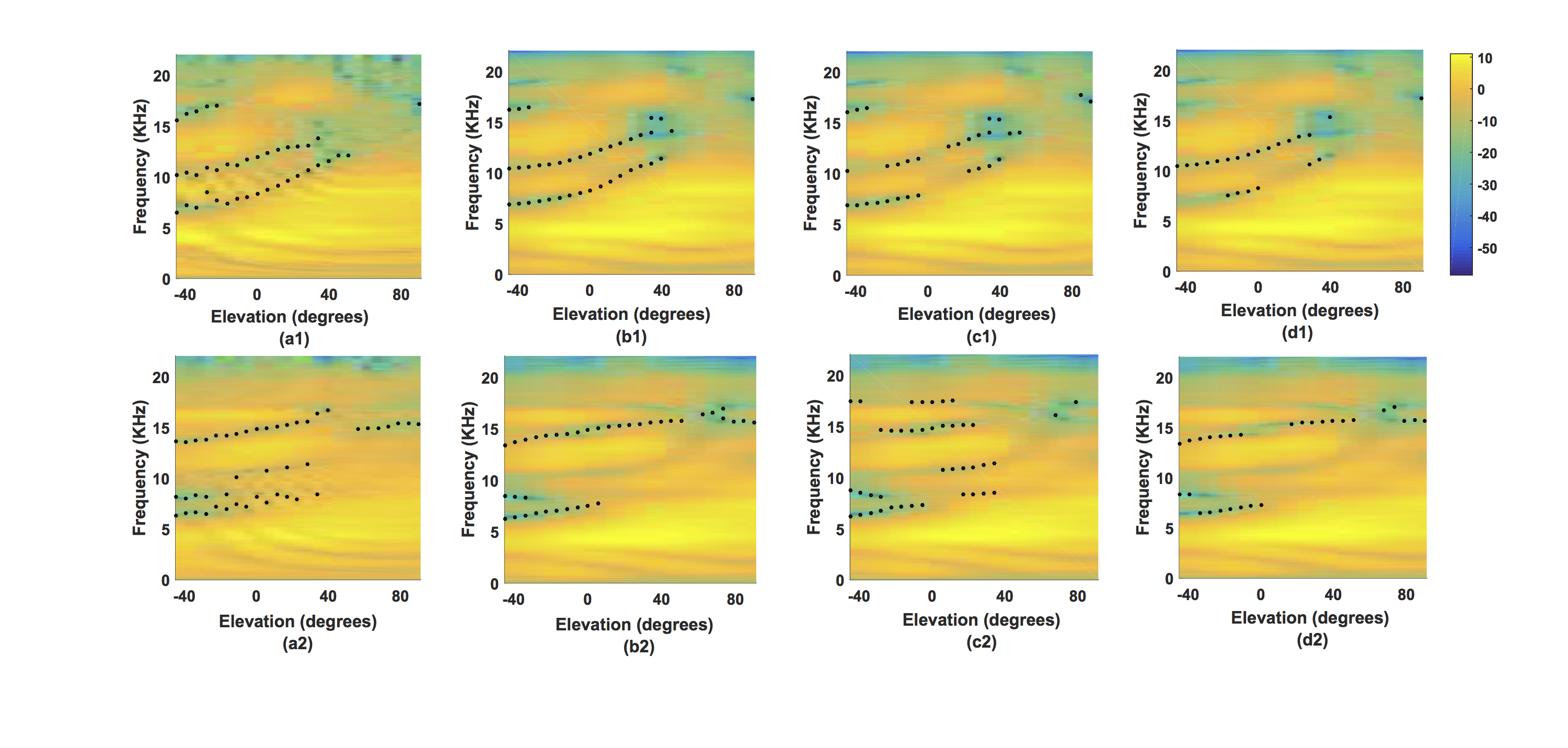}}
\vspace{-1cm}
   \caption{Variation of PSN frequencies with elevation angle for measured HRTF (a1,a2), proposed M-HRTF model HRTF (b1,b2), Minimum phase component of HRTF (c1,c2), and all pass component of HRTF (d1,d2) corresponding to  Subjects (163, 119) respectively of CIPIC database}
   \label{fig:notchplot119-163}         
\end{center}
%\hfill
\end{figure*}

\begin{figure*}
\centerline{\includegraphics[width=18cm, height=9cm,trim=10 10 10 10,clip]{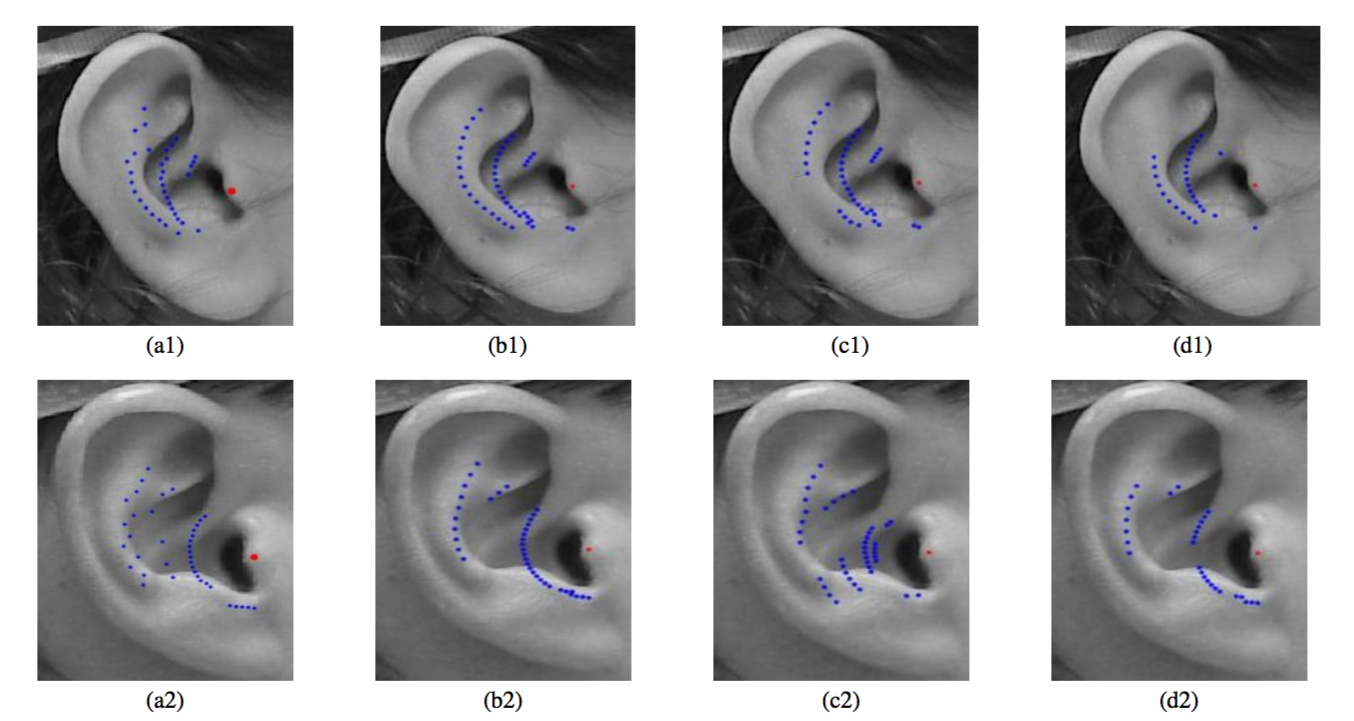}}
\caption{Pinna contours obtained from the PSNs extracted for measured HRIR (a1,a2), reconstructed HRIR using proposed method  (b1,b2), minimum phase component of reconstructed HRIR (c1,c2), all pass component of reconstructed HRIR (d1,d2)  are mapped to the pinna images available on CIPIC database for subjects 163, and 119 respectively.}
\label{fig:Mymapping}
\end{figure*}

\vspace{-0.5cm}

\section{Performance Evaluation}
\label{sec:4}

Performance evaluation of the proposed HRTF model is evaluated by extracting and investigating the quality of PSNs obtained from the HRIR. Coherence analysis between HRIR computed from the proposed model and the ground truth HRIR is performed. Finally  binaural audio is synthesised using the HRIRs obtained herein. Subjective evaluations of the rendered binaural audio is performed and compared with HRIRs obtained from CIPIC, KEMAR and MiPS databases.

\vspace{-0.3cm}

\subsection{Databases Used}
\subsubsection{CIPIC Database}
\vspace{-0.1cm}
 CIPIC database \cite{cipic} contains HRIRs measured by following interaural polar coordinate system. CIPIC contains total of 1250 HRIRs with 25 lateral angles ranging from $-80^0$ to $80^0$ and 50 elevation angles ranging from $-45^0$ to $230.625^0$. There are 50 HRIRs in both horizontal and median plane. Each HRIR is of 200 sample length and measured at a sampling frequency of 44100 Hz. In addition to HRIRs, CIPIC also includes left and right pinna images of subjects for whom the HRIR measurement has been performed. 

\subsubsection{KEMAR Database}
KEMAR database \cite{gardner} contains HRIRs measured by following vertical polar coordinate system . KEMAR contains HRIRs  for elevations angles ranging from $-40^0$ to $90^0$ with an angular increment of $10^0$, and for azimuthal angles ranging from $-170^0$ to $180^0$ with an angular increment of $5^0$. KEMAR contains 28 HRIRs in median plane, and 72 HRIRs in the horizontal plane. Each HRIR is of 512 sample length and measured at a sampling frequency of 44100 Hz.

\subsubsection{MiPS Database}
MiPS database  contains HRIRs measured for the sampling points  chosen by combining the interaural and vertical polar coordinate as discussed in \cite{bionic}. HRIRs are measured for every $10^0$ resolution in the vertical circles. MiPS database contains 36 HRIRs in both horizontal and median plane. Each HRIR is of 1024 sample length and measured at a sampling frequency of 44100 Hz.

\subsection{Experiments on Median Plane PSN Extraction}
 As the proposed HRTF model is aimed at preserving the important spectral cues, pinna spectral notches are extracted for the HRTFs obtained from the proposed model and compared it with the PSNs of measured HRTFs and the PSNs of minimum phase-pure delay model. Further we also illustrate the spectral notches of the all pass components in order to illustrate the spectral notches that are lost due to the minimum phase approximation of HRTFs.  In this work, PSNs are extracted using LP-GD method with a threshold for notch depth equal to -0.8. PSNs extracted for composite HRTFs (a1,a2), reconstructed HRTFs using proposed method (b1,b2), minimum phase component of composite HRTF (c1,c2), and all pass component of composite HRTF (d1,d2) in the median plane for elevations $-45^0$ to $90^0$ are illustrated in Fig. \ref{fig:notchplot119-163}. As compared to Fig. \ref{fig:notchplot119-163} (a1) and (a2), the spectral notches are smoother  in Fig. \ref{fig:notchplot119-163} (b1) and (b2). This can be attributed to the FBS method of HRTF reconstruction, since this method facilitates continuous evolution of spectral notches. PSNs are observed to be discontinuous in both Fig. \ref{fig:notchplot119-163}(c1),(c2) and (d1),(d2). This indicates that both minimum phase and all pass components do not exhibit well defined spectral notches for all elevation angles. Additionally it must be noted that, not all the directions in the median plane are pure minimum phase. PSNs extracted for the proposed HRTFs are smoother and evolve continuously. This smooth evolution observed in the PSN can be noted when PSNs are mapped to pinna images. 

\subsubsection{Mapping Pinna Spectral Notches to Pinna Images}

 The relation between PSNs and contours of the pinna has been studied  using a two ray reflection model \cite{raykar,PRTF}. A similar mapping is performed for the composite HRTFs, reconstructed HRTFs using proposed method,  minimum phase and all pass components  of composite HRTFs. It can be observed from Fig. \ref{fig:Mymapping} that, notch distances vary smoothly for the reconstructed HRTFs as compared to distances obtained for composite HRTFs.  Additionally in the proposed method, notch distances can be obtained for all the elevation angles in a seamless manner. An important observation in the mappings of minimum phase and all pass components is that, some of the spectral notches are unique to either minimum phase or all pass components. For example in Fig. \ref{fig:Mymapping}(c1) some of the PSN mappings are missing in the middle of concha wall, where as in Fig. \ref{fig:Mymapping}(d1), those PSN mappings can be observed. This indicates that, in some regions spectral notches contributed by the all pass components are directly related to the sound reflections from the concha wall. On the other hand the PSN mappings exhibited at the top of concha wall in Fig. \ref{fig:Mymapping}(c1), is absent in Fig. \ref{fig:Mymapping}(d1). These regions exhibit purely minimum phase behaviour of HRTFs. A similar observation can be made for Subject 163 of CIPIC database in Figs. \ref{fig:Mymapping}(a2)-(d2). This indicates that ignoring all pass spectral notches would prune some of the sound reflections from the the concha wall for directions which do not exhibit purely minimum phase behaviour.

\begin{figure*}[b!]
\begin{center}
   \centering
\centerline{\includegraphics[width=20cm, height=12cm,trim=110 10 120 10,clip]{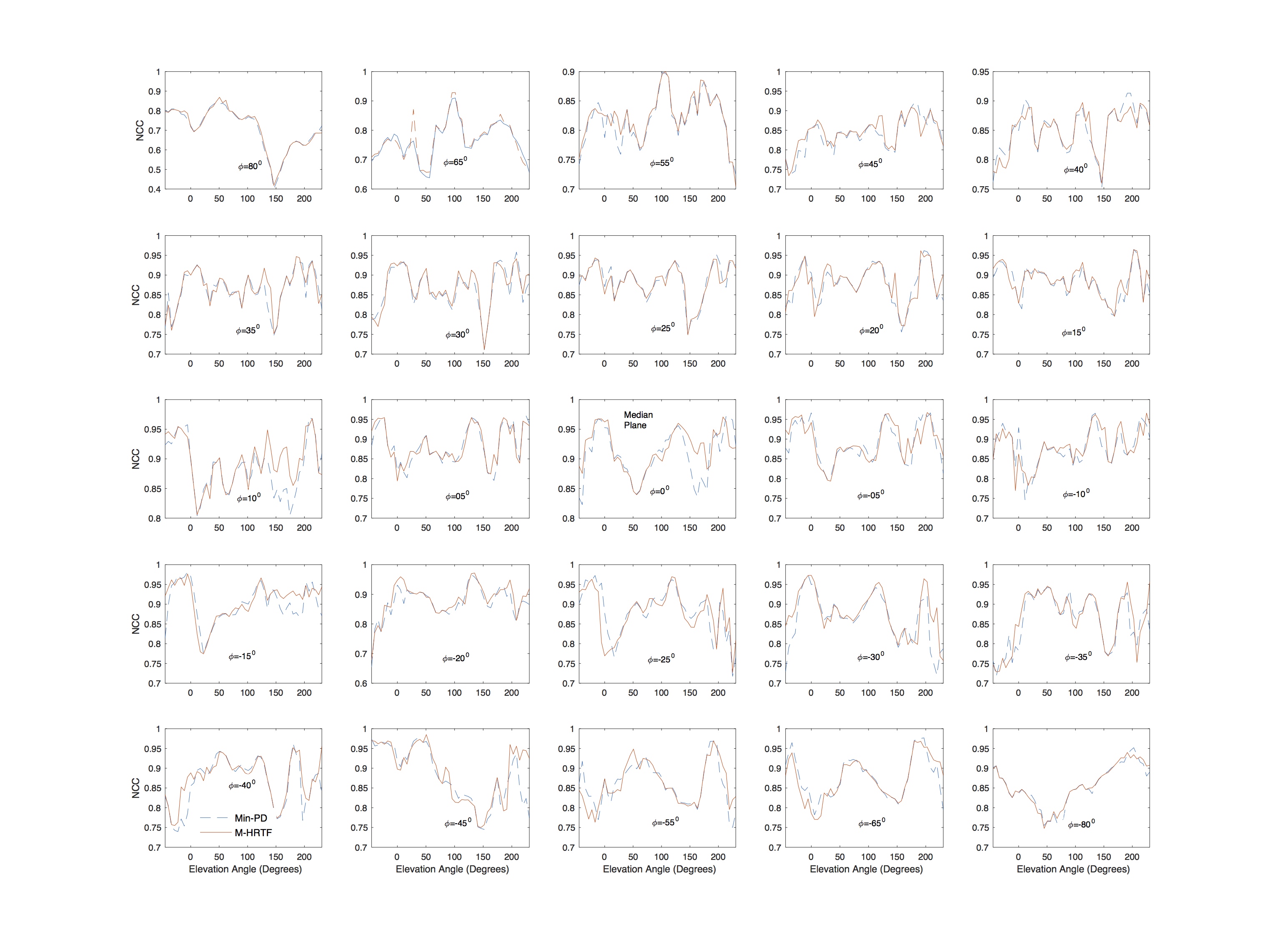}}
\vspace{-0.3cm}
   \caption{Normalized Cross Coherence (NCC) between HRIR for various elevation and azimuthal angles of CIPIC database. NCC between composite HRIR and HRIR obtained using Min-PD model is indicated in Blue. NCC between composite HRIR and HRIR obtained using proposed M-HRTF model is indicated in Red.}
   \label{fig:coherence_complete}         
\end{center}
\end{figure*}
%
%

%\vspace{-0.5cm}

\vspace{-0.2cm}
\subsection{Cross Coherence Analysis}

\vspace{-0.2cm}

\begin{figure}
\begin{center}
   \centering
\centerline{\includegraphics[width=9cm, height=5.5cm,trim=50 10 50 0,clip]{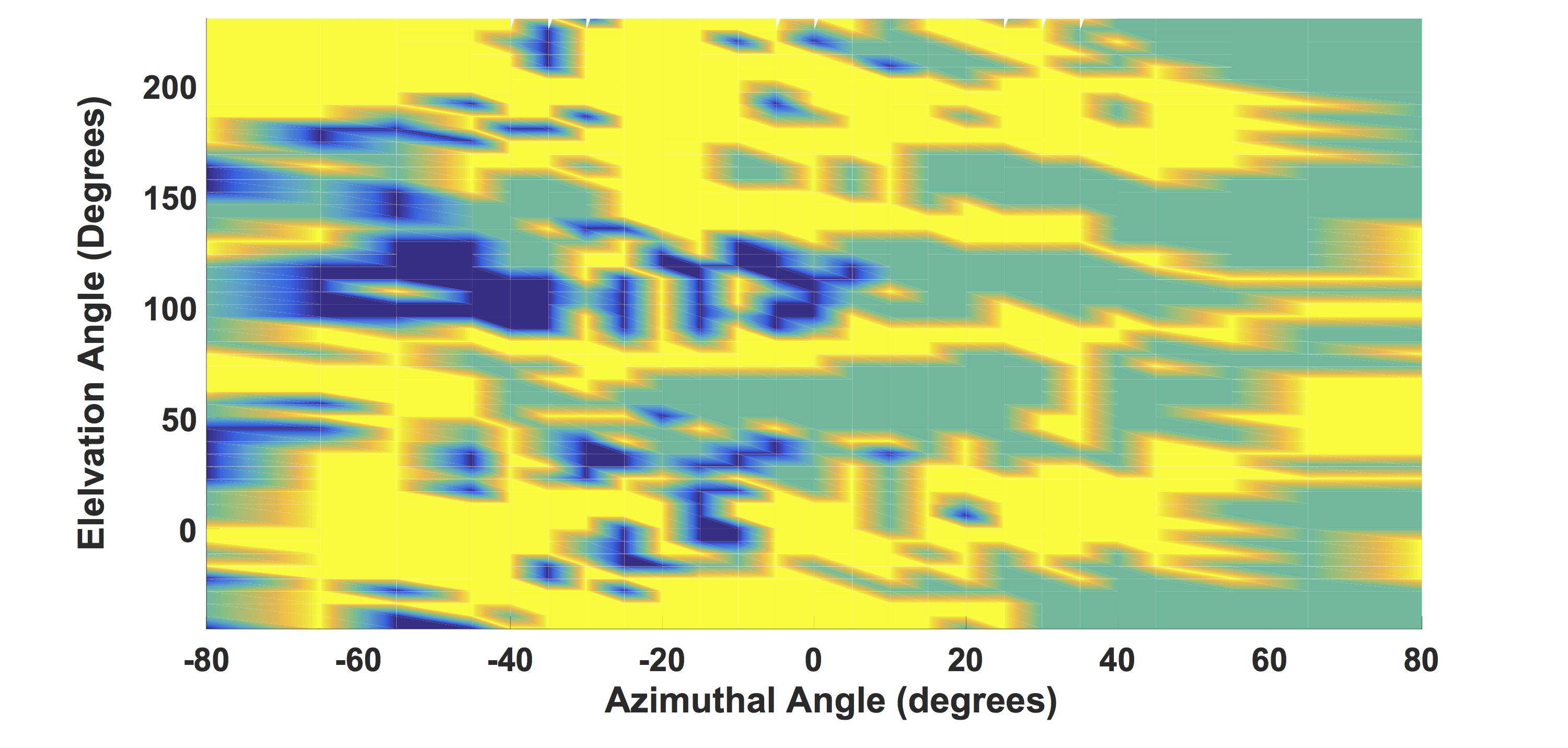}}
%\vspace{-0.3cm}
   \caption{Variation of $(\psi^{*}_{D})$ for different elevation and azimuthal angles.  Yellow regions indicate that the proposed M-HRTF model yields high cross coherence for majority of azimuthal and elevation angles.}
   \label{fig:coherence_bar}         
\end{center}
%\hfill
\end{figure}

Minimum phase-pure delay model (Min-PD) ignores the phase contributed by the all pass component. The proposed HRTF model (M-HRTF) captures the contribution of  both the pure minimum phase and minimum phase-all pass behaviour. In order to evaluate both these models, an objective measure called Normalised Cross Coherence (NCC) \cite{stanford} is used herein. The expression for NCC is given by

\begin{equation}
\psi_{i}(k)=\frac{\sum\limits_{n=0}^{\infty} h_{gt}[n-k]h_{i}[n]}{{\bigg[\sum\limits_{n=0}^{\infty} h_{gt}^2[n].\sum\limits_{n=0}^{\infty} h_{i}^2[n]\bigg]}^{\frac{1}{2}}}
\label{coherence}
\end{equation}

%\vspace{-0.3cm}

\begin{equation}
\psi^{*}_{i}=\max_{k}\{\psi_{i}(k)\} \quad i \in \{apf,mpd\}
\end{equation}
where, $h_{gt}$, $h_{apf}$, $h_{mpd}$ indicate the ground truth HRIR, HRIR obtained from M-HRTF model, HRIR obtained from Min-PD model respectively. However there exists an onset time between minimum phase and composite HRIRs. Therefore the point at which there is a maximum coherence $(\psi^{*})$ can be used as a measure to determine correlation between composite HRIRs and the HRIRs reconstructed from different models.  HRIR cross coherence computed for all the directions of CIPIC database are illustrated in Fig.  \ref{fig:coherence_complete}.  It can be observed that, for elevation angles ranging from $50^0$ to $120^0$ the cross coherence is same for the proposed M-HRTF model and the Min-PD model.  It is so because, these directions exhibit purely minimum phase behaviour. For the remaining directions we can observe a higher cross coherence for the M-HRTF model when compared to Min-PD model for most of the directions. Difference in the cross coherence is used as a measure to identify the model which exhibits higher coherence. This difference in normalised cross coherence can be defined as,
\begin{equation}
\psi^{*}_{D}=\psi_{apf}(k)-\psi_{mpd}(k)
\end{equation} 
$\psi^{*}_{D}$ indicate the difference in the cross coherence obtained by M-HRTF and Min-PD model.  A single dimensional K-means clustering is used to segregate the cross coherence values in to three categories.  In Fig. \ref{fig:coherence_bar}, the first category as shown in color (Yellow) indicate HRIRs of these directions exhibit higher cross coherence for M-HRTF model. The second category as shown in color (Blue) indicate HRIRs of these directions exhibit the higher coherence for Min-PD model, and the third category as shown in color (green) indicate the nearly similar cross coherence with respect to HRIRs of both models. It can be observed that HRIRs obtained using M-HRTF model has higher coherence for a larger number of directions as compared to HRIRs obtained from Min-PD model.

%\begin{figure}
%\begin{center}
%   \centering
%\centerline{\includegraphics[width=10cm, height=6cm,trim=120 50 40 45,clip]{NormalizedCoherence_MP1_f}}
%\vspace{0.4cm}
%   \caption{Normalised Cross Coherence (NCC) between HRIR for various elevation angles in the Median plane of CIPIC database. NCC between composite HRIR and HRIR obtained using Min-PD model is indicated in Blue. NCC between composite HRIR and HRIR obtained using proposed M-HRTF model is indicated in Red.}
%   \label{fig:coherence_back}         
%\end{center}
%\end{figure}

%\vspace{-0.5cm}
 
 \subsection{Experiments on  Binaural Sound Synthesis}
 HRTFs obtained from the proposed method are utilised for  binaural sound synthesis.  As median plane HRTFs are highly person specific, binaural sound synthesis is performed only in the horizontal plane. FBS method is used to reconstruct HRTFs for any angle in the 3D space. HRTFs are obtained by proposed method and Min-PD model for azimuthal angles with a resolution of $1^0$. Measured HRTFs of CIPIC, KEMAR, and MiPS databases in the horizontal plane are also used for evaluations. A monaural sound signal is spatialised using HRTFs obtained from various methods. Binaural rendering is performed by computing
 block convolution, where, successive blocks of
monophonic sound are convolved with HRIRs of successive
directions in the 3D space. Subsequently these filtered sounds
are concatenated using the overlap save method \cite{oppenheim2010discrete}. The filtered signal using left and right HRIRs are played through headphones. The synthesised binaural audio is evaluated subjectively using Mean Opinion Scores (MOS). Fifteen subjects are selected in the evaluation. Following attributes are used for subjective evaluations \cite{sub_eval_3,SAQI}.

\begin{itemize}%[\setlabelwidth{12)}]
\item Naturalness (N): How true to life the audio listening was?
\vspace{0.2cm}
\item Presence (Ps): Presence in audio source environment.
\vspace{0.2cm}
\item Preference (Pf): Degree of pleasantness or harshness.
\vspace{0.2cm}
\item Perception of Motion (PoM): The precision and correctness with which the trajectory of the source is perceived.
\end{itemize}
The corresponding MOS obtained for spatial sound  are enumerated in Table \ref{tablesub1}.
The difference between the MOS of PoM attribute across various methods is very low. But measured HRTFs and proposed method HRTFs has slightly better MOS when compared to MOS obtained for Min-PD HRTFs. The MOS of preference attribute is found to have higher value and is mostly same for all the methods. For naturalness attribute, measured HRTFs seem to provide better quality of rendered sound. In the presence attribute, measured HRTFs followed by proposed method HRTFs seem to have higher scores. Across all the attributes measured HRTF and proposed method HRTFs has higher scores when compared to the HRTFs obtained by the Min-PD model. 

\begin{table}
\caption{Mean Opinion Scores obtained for  binaural  synthesis in Horizontal plane}
\begin{center}
\begin{tabular}{|l|l|l|l|l|}
\hline
 & \textbf{N} & \textbf{Ps} & \textbf{Pf}  & \textbf{PoM} \\ \hline
\textbf{CIPIC-Measured} & 4.1 & 4.2 & 4.2  & 4 \\ \hline
\textbf{CIPIC-Min-PD} & 3.8 & 3.8 & 4.2  & 3.5 \\ \hline
\textbf{CIPIC-Proposed} & 3.8 & 4.1 & 4.1  & 3.9 \\ \hline
\textbf{KEMAR-Measured} & 3.8 & 4 & 4.2  & 3.9 \\ \hline
\textbf{KEMAR-Min-PD} & 3.6 & 3.8 & 4.2  & 3.5 \\ \hline
\textbf{KEMAR-Proposed} & 3.6 & 4 & 4  & 3.8 \\ \hline
\textbf{MiPS-Measured} & 3.5 & 4.2 & 4.1  & 3.7 \\ \hline
\textbf{MiPS-Min-PD} & 3.5 & 3.7 & 4  & 3.3 \\ \hline
\textbf{MiPS-Proposed} & 3.5 & 3.7 & 4.1  & 3.7 \\ \hline
\end{tabular}
\end{center}
\label{tablesub1}
\end{table}

%\vspace{0.3cm}

\section{Conclusion}
\label{finalsec}
A new HRTF model is developed in this paper for accurate
reconstruction of HRTFs. A novel group delay decomposition method is used
in developing the proposed model which captures the pinna spectral
notches due to both the minimum phase and all pass component of the HRTF.
Selectively including the all pass filter for specific angles in to the proposed model improved the accuracy of capturing the HRTF phase response. Additional advantages of the proposed HRTF reconstruction model include a
PSN extraction with a high degree of resolution and reduced cost of the binaural sound synthesis without compromising the important spectral cues. Future work will investigate the computational complexity of the proposed  HRTF model in  binaural sound synthesis applications. Development of real time binaural sound systems using the proposed HRTF model will also be investigated in future.

%\vspace{-0.5cm}

\section{Acknowledgments}
This work was funded in part by TCS Research Scholarship Program under project number TCS/CS/2011191E and in part by SERB, Dept. of Science and Technology, GoI under project number SERB/EE/2017242.


\begin{thebibliography}{99}

\bibitem{begault20003}  Begault  D. R. and  Trejo L. J.: '3-d sound for virtual reality and multimedia', 2000.
\bibitem{blauert} Blauert, J.: 'Spatial hearing: the psychophysics of human sound localization', Cambridge, Mass. MIT Press, 1997. [Online]. Available: http://opac.inria.fr/record=b1126831.
\bibitem{cipic} Algazi, V., Duda, R., Thompson, D., and Avendano, C.: 'The cipic hrtf database', in IEEE Workshop on the Applications of Signal Processing to Audio and Acoustics, 2001, pp. 99-102.
\bibitem{6600896} Jin, C. T.,  Guillon, P., Epain,N., Zolfaghari, R.,  Schaik, A.V.,  Tew, A. I.,  Hetherington, C., and  Thorpe, J.: ' Creating the
sydney york morphological and acoustic recordings of ears database', in IEEE Transactions on Multimedia, vol. 16, no. 1, pp. 37?46, Jan 2014.
\bibitem{kulkarni} Kulkarni, A., Isabelle, S., and Colburn, H.: 'On the minimum-phase approximation of head-related transfer functions' , in IEEE ASSP Workshop on Applications of Signal Processing to Audio and Acoustics, 1995, pp. 84-87.
\bibitem{stanford} Nam, J., Kolar, M., and Abel,  J.: 'On the minimum-phase nature of head-related transfer functions', in the 125th Audio Engineering Society Convention, AES. San Francisco, CA, USA: AES, 2008. [Online]. Available: http://www.aes.org/e-lib/browse.cfm?elib=14698.
\bibitem{Wright} Wright, D., Hebrank, J. H.,  and Wilson, B.: 'Pinna reflections as cues for localization', The Journal of the Acoustical Society of America, vol. 56, no. 3, 1974.
\bibitem{karan_MM} Nathwani, K., Pandit, P., and Hegde, R. M.:  'Group delay based methods for speaker segregation and its application in multimedia information retrieval', IEEE Transactions on Multimedia, vol. 15, no. 6, pp. 1326-1339, Oct 2013.
\bibitem{raykar}  Raykar, V. C., Duraiswami, R.,  and Yegnanarayana, B.: 'Extracting the frequencies of the pinna spectral notches in measured head related impulse responses', The Journal of the Acoustical Society of America, vol. 118, no. 1, 2005.
\bibitem{abhayahorz} Abhayapala, T. D.,  Kennedy, R. A.  and Zhang, W.: 'Horizontal plane hrtf reproduction using continuous fourier-bessel functions', in Audio Engineering Society Conference: 31st International Conference: New Directions in High Resolution Audio, Jun 2007. [Online]. Available: http://www.aes.org/e-lib/browse.cfm?elib=13969
\bibitem{FILTERS07} Smith, J. O.:  'Introduction to Digital Filters with Audio Applications', W3K Publishing, 2007. [Online]. Available: http://www.w3k.org/books/
\bibitem{FILTERSWEB07} Smith, J. O.:  'Introduction to digital filters with audio applications' , Available  [Online]: http://ccrma.stanford.edu/ jos/filters.
\bibitem{oppenheim2010discrete} Oppenheim A. V.,  and Schafer, R. W., 'Discrete-time signal processing', Pearson Higher Education, 2010.
\bibitem{yegnanarayana1992significance} Yegnanarayana, B., and Murthy, H.A.: 'Significance of group delay functions in spectrum estimation', IEEE Transactionson signal processing, vol. 40, no. 9, pp. 2281-2289, 1992. [Online]. Available: http://dx.doi.org/10.1007/s12046-011-0045-1
\bibitem{kumar2014robust} Kumar, L.,  Tripathy, A.,  and Hegde, R. M.:  'Robust multi-source localization over planar arrays using music-group delay spectrum', IEEE Transactions on Signal Processing, vol. 62, no. 17, pp. 4627-4636, 2014.
\bibitem{4032772} Hegde, R. M., Murthy, H. A., and  Gadde, V. R. R. 'Significance of the modified group delay feature in speech recognition', IEEE Transactions on Audio, Speech, and Language Processing, vol. 15, no. 1, pp. 190-202, Jan 2007.
\bibitem{abhaya} Zhang, W.,  Abhayapala, T. D., Kennedy, R. A., and Duraiswami, R. : 'Insights into head-related transfer function: Spatial dimensionality and continuous representation', The Journal of the Acoustical Society of America, vol. 127, no. 4, 2010.
\bibitem{gardner} Gardner, B.  Martin K.  et al.: 'Hrtf measurements of a kemar dummy-head microphone', Massachusetts Institute of Technology, vol. 280, no. 280, pp. 1-7, 1994.
\bibitem{bionic} Reddy S.  and Hegde, R. M.:  'Design and development of bionic ears for rendering binaural audio', in 2016 International Conference on Signal Processing and Communications (SPCOM), 2016.
\bibitem{PRTF} Spagnol, S.,  Geronazzo, M.,  and Avanzini, F., 'Fitting pinna-related transfer functions to anthropometry for binaural sound rendering', in IEEE International Workshop on Multimedia Signal Processing (MMSP),  Oct 2010, pp. 194-199.
\bibitem{sub_eval_3} Guastavino C.,  and Katz, B. F. G.:  'Perceptual evaluation of multi-dimensional spatial audio reproduction', The Journal of the Acoustical Society of America, vol. 116, no. 2, 2004.
\bibitem{SAQI} Lindau, A.,  Erbes, V.,  Lepa, S.,  Maempel, H.J.,  Brinkman, F.,  and Weinzierl, S.: 'A spatial audio quality inventory (saqi)', Acta Acustica united with Acustica, vol. 100, no. 5, 2014.

\end{thebibliography}
\end{document}